\documentclass[review]{elsarticle}

\usepackage{lineno}

\usepackage{geometry}

\usepackage{subfigure}
\usepackage{amsmath}
\usepackage{amssymb}
\usepackage{color}
\usepackage{graphicx}
\usepackage{url}
\usepackage{graphicx}
\usepackage{dcolumn}
\usepackage{bm}
\usepackage{comment} 
\usepackage{multirow}
\usepackage{amsthm}
\usepackage{algorithm}
\usepackage{algorithmic}
\usepackage{pseudocode}

\usepackage[english]{babel}
\makeatletter
\adddialect\l@en\l@english
\makeatother

\usepackage{xcolor}

\begin{document}

\begin{frontmatter}
\title{Multidimensional epidemic thresholds \\ in diffusion processes over interdependent networks\tnoteref{mytitlenote}}
\tnotetext[mytitlenote]{This work has been partly funded by FIRB project Information monitoring, propagation analysis and community detection in Social Network Sites.}

\author[mymainaddress,mysecondaryaddress]{Mostafa Salehi\corref{mycorrespondingauthor}}
\cortext[mycorrespondingauthor]{Corresponding author (mostafa.salehi@unibo.it)}
\address[mymainaddress]{University of Bologna, Italy}
\address[mysecondaryaddress]{University of Tehran, Iran}
\ead{mostafa.salehi@unibo.it}

\author[Payammainaddress]{Payam Siyari}
\address[Payammainaddress]{Sharif University of Technology, Iran}
\ead{siyari@ce.sharif.edu}

\author[Matteomainaddress]{Matteo Magnani}
\address[Matteomainaddress]{Uppsala University, Sweden}
\ead{matteo.magnani@it.uu.se}

\author[mymainaddress]{Danilo Montesi}
\ead{danilo.montesi@unibo.it}

\begin{abstract}
Several systems can be modeled as sets of interdependent networks where each network contains distinct nodes. 
Diffusion processes like the spreading of a disease or the propagation of information constitute fundamental phenomena occurring over such coupled networks. In this paper we propose a new concept of \emph{multidimensional epidemic threshold} characterizing diffusion processes over interdependent networks, allowing different diffusion rates on the different networks and arbitrary degree distributions. We analytically derive and numerically illustrate the conditions for multilayer epidemics, i.e., the appearance of a giant connected component spanning all the networks. Furthermore, we study the evolution of infection density and diffusion dynamics with extensive simulation experiments on synthetic and real networks.
\end{abstract}


\end{frontmatter}


\section{Introduction}\label{sec:intro}

Information diffusion events like the spreading of rumours and behaviors or the coverage of news headlines through different weblogs represent an important class of dynamical processes on networks.
However, while the study of diffusion processes over single networks has received a great deal of interest from various disciplines for over a decade \cite{Borge-Holthoefer2013a}, real diffusion phenomena are seldom constrained inside a single network. A typical example is represented by the diffusion of epidemics propagated by human beings traveling via multiple transport networks (airplanes, trains, etc.). Therefore, a great deal of interest has been recently devoted to the study of diffusion processes on \emph{multilayer networks} \cite{DBLP:conf/asonam/MagnaniR11,Kivela2013,Boccaletti2014,Salehi2014}.

While most works on diffusion in multilayer networks have focused on a specific type of network often called \emph{multiplex}, where the same nodes are present on all networks \cite{Min2013,Yagan2013,Qian2012,Qian2013}, less attention has been devoted to diffusion processes on systems of interdependent networks with distinct nodes on the different layers. For brevity, we will refer to these as \emph{disjoint interdependent networks}. To the best of our knowledge, the only work focusing on this kind of multilayer networks is \cite{Dickison2012a}. Specifically, in this work it has been shown that in the case of diffusion over pairs of interconnected Erd\H{o}s-Renyi networks a mixed phase can occur, where the diffused items are mainly spread only on one of the networks.

This result is interesting because it highlights the roles of the internal layer structures and of their interconnections in diffusion processes. However, it is based on some strong assumptions: networks are assumed to belong to the Erd\H{o}s-Renyi family, which is rarely observed in real cases, and a single diffusion rate for all inter- and intra-layer connections is set. As an example reusing a motivating scenario introduced in \cite{Dickison2012a}, this last assumption would state that the diffusion of a disease between two people living in the same city (here modeled as a citizen network) has the same probability to be transmitted between two people from different cities.

In this work we relax these assumptions and provide a more general treatment of the problem of information diffusion over disjoint interdependent networks, allowing for different network structures and different diffusion rates for inter- and intra-layer connections. We introduce the concept of multidimensional epidemic threshold and analytically extract the conditions of epidemics mapping the original network to a colored degree-driven random graph (CDRG) \cite{Qian2012}. We then present simulation results on synthetic and real interdependent networks to validate our analytical findings and to study the behavior of other variables typically used to describe diffusion processes.

\section{Preliminaries and Definitions}\label{sec:pre}

\begin{figure}[!tp]
\begin{center}
\includegraphics[trim=30mm 50mm 10mm 30mm, scale=0.35]{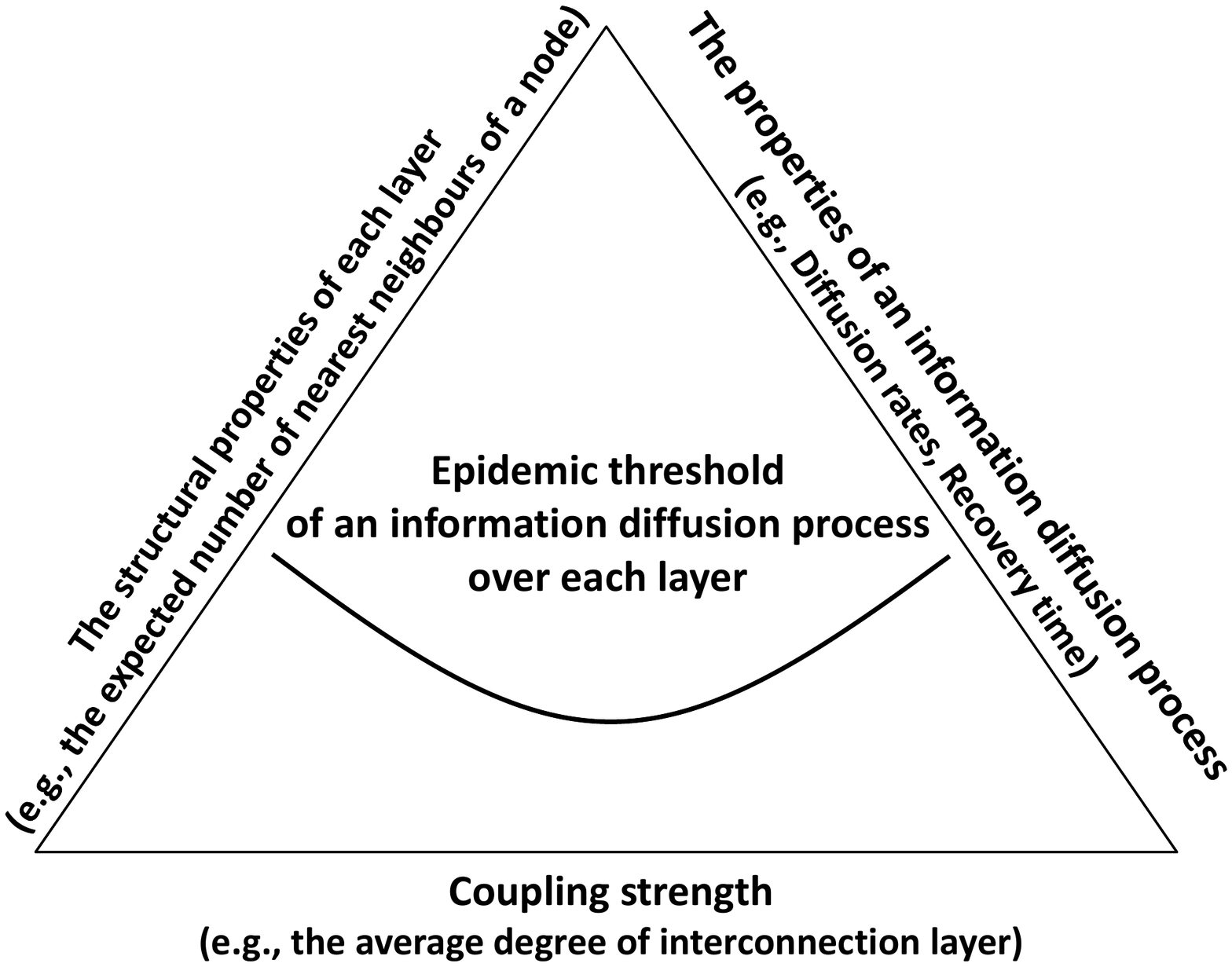}
\end{center}
\caption{The main factors in the dynamics of information diffusion over interdependent networks.}
\label{fig:fig-epiTh3}
\end{figure}

Figure \ref{fig:fig-epiTh3} summarizes the main factors governing a diffusion process across a set of disjoint interdependent networks. The condition for the occurrence of an epidemic is influenced by the interplay between the general properties of the information diffusion process (i.e., the diffusion model and its parameters), the structural properties of each network and the coupling strength between pairs of networks.

In this article we use SIR (susceptible-infected-recovered) as our general model of diffusion, following several other works on the modeling of diseases and influence on multilayer networks \cite{Min2013,Yagan2013,Qian2012,Dickison2012a,Wang2011a,Zhao2014770,Marceau2011,Cozzo2013a,PhysRevE.81.036118}. 
As a very quick reminder of this model, in SIR it is assumed that the diffusion process starts with an initial set of infected\footnote{Terms like infection and epidemic are normally used to refer to a wide range of processes with similar macroscopic behaviors, including information diffusion. While we are critical towards the indiscriminate usage of this terminology to describe general diffusion processes not involving viruses, we adopt it in this paper to simplify references to the literature.} nodes called seeds. An infected node can propagate its status to a susceptible neighbor with diffusion rate $\beta$.
In the version of the SIR model used in this article, infected nodes recover after time $\tau$ from the moment of infection. 
The transmissibility of the network, i.e., the probability that an infected node propagates to a susceptible neighbor before recovery time, can be computed as $R=1-{(1- \beta)}^\tau$ \cite{Newman2002}.
 
The aforementioned parameters can be utilized to compute the epidemic threshold \cite{PhysRevE.81.036118}, which is one of the key values used to characterize a diffusion process. This indicates a value of infection rate above which the diffusion network (i.e., the actual nodes and links traversed during the diffusion process) constitutes a giant connected component (gcc) with respect to the underlying network. By definition, a gcc contains a finite fraction of nodes in the limit of large network sizes.
 
Let $\kappa$ be the average degree of a randomly chosen end vertex of a randomly chosen edge~\cite{Dorogovtsev2003}.
To diffuse information over all nodes in an arbitrary (connected) network, each infected node must infect on average at least one of its neighbors. Then, for large single networks the epidemic threshold can be computed as \cite{Dickison2012a}:
\begin{equation}\label{eq:epiTheSingle}
\beta_c = 1-[1-(\kappa -1)^{-1}]^{1/\tau}
\end{equation}

Eq.~(\ref{eq:epiTheSingle}) is based on the assumption that the diffusion rate is the same for all links in the underlying network. Therefore, assuming the same diffusion rate $\beta$ for all the networks and for all their interconnections one can directly utilize this equation to compute the epidemic threshold for an arbitrary interdependent network, where $\kappa$ is calculated over the entire coupled network (i.e., including all intra- and inter-layer links) \cite{Dickison2012a}. 
When different diffusion rates are considered for inter- and intra-layer links, this equation can no longer be used and the concept of a single threshold no longer captures the complexity of the process.

\begin{figure}[!tp]
\begin{center}
\includegraphics[width=\textwidth]{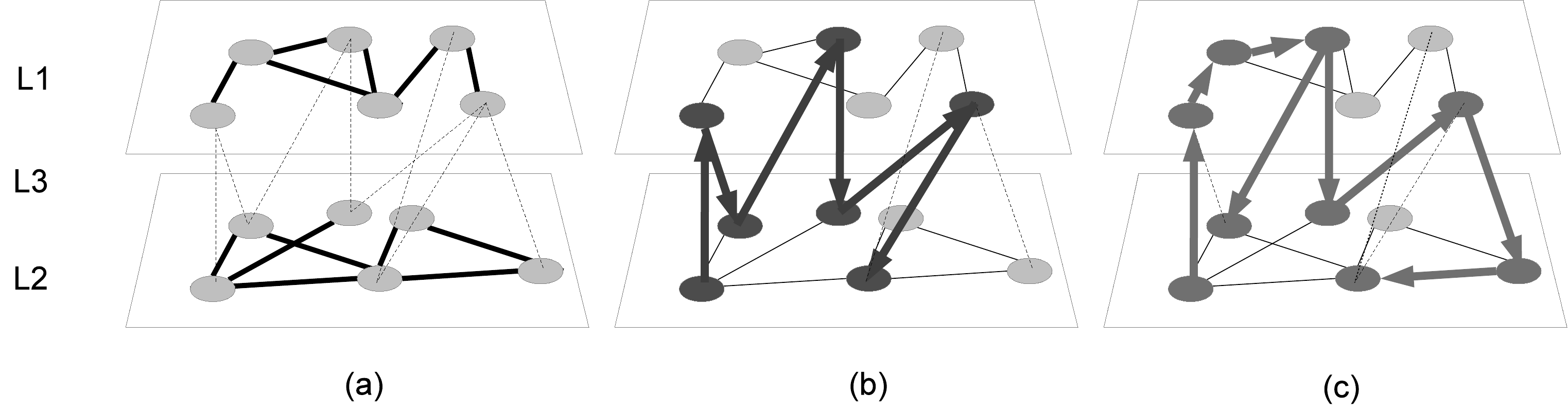}
\end{center}
\caption{
(a) An interdependent network with two disjoint layers ($L_1$ and $L_2$) and an interconnection layer ($L_3$), (b) an epidemic spreading through inter-layer links and (c) a different diffusion process exploiting a combination of intra- and inter-layer links.
}

\label{fig:kindsof}
\end{figure}

As an example, consider Figure \ref{fig:kindsof}(a): we have two networks $L_1$ and $L_2$ with interconnection layer $L_3$ (that is, $L_3$ contains connections between nodes in $L_1$ and nodes in $L_2$). Let us call the diffusion rates on $L_1$, $L_2$ and $L_3$ respectively $\beta_1$, $\beta_2$ and $\beta_3$. Then, it is worth noticing that there is no longer a single possible threshold. For example, if the interconnection layer $L_3$ is characterized by a high infection rate, a giant connected component can be generated just having the diffusion happening on the inter-layer links, with nodes in $L_1$  iteratively infecting nodes on $L_2$ and viceversa (Figure \ref{fig:kindsof}(b)). However, even in case of a lower inter-layer diffusion rate that would not be enough alone to generate an epidemic, a giant component could still emerge thanks to additional infections inside $L_1$ and $L_2$ (Figure \ref{fig:kindsof}(c)). As an example, the two tuples (0, 0, 0.22) and (0.02, 0.02, 0.15) can both represent values above which $\beta_1, \beta_2, \beta_3$ generate an epidemic, while no epidemic might occur with $(\beta_1, \beta_2, \beta_3)$ = (0.01, 0.01, 0.16).

Following these considerations we introduce the concept of multidimensional threshold: 

\textbf{Dominance:} 
A tuple $t_1$ in $[0,1]^n$ dominates a tuple $t_2$ if 
$\forall i \in [1, n]: t_1(i) \leq t_2(i) \land \exists l: t_1(l) < t_2(l)$. We then write $t_1 \succ t_2$

\textbf{Multidimensional epidemic threshold:}
Let $0 \leqslant \beta_i \leqslant 1, i \in [1,n]$ be diffusion rates on single layers or between two layers, and $N = [0,1]^n$ be the set of all possible tuples.
The multidimensional epidemic threshold of information diffusion over an interdependent network is a set $M \subset N$ such that 
(i) $\forall t' \in N, t \in M, t \succ t'$ a gcc on the whole network is generated with high probability and 
(ii) $\forall t \in M \ \nexists t' \in N$ such that $t'$ both dominates $t$ and determines the emergence of a gcc.

In terms of the amount of information spreading over the entire network and its layers, one can define three different states for an interdependent network \cite{Dickison2012a}: (i) infection-free; where there is no epidemic in any of the layers (i.e., there is no gcc), (ii) mixed; where a gcc appears only in some of the layers, and (iii) epidemic; where the gcc appears in all layers.
\section{Analytical Results}\label{sec:anal}

In this section we analytically study the relation between epidemic threshold and diffusion rates in the general case of disjoint interdependent networks with different intra- and inter-layer rates.

To this end, we map our problem into the framework of inhomogeneous random graphs, also called colored degree-driven random graphs (CDRG) \cite{Soderberg2003a, Soderberg2003b, Soderberg2003c}. Recently, the CDRG framework was used in \cite{Qian2012} to study information diffusion over multiplex networks. Here we extend it to an arbitrary set of interdependent networks.

In this framework each edge between two vertexes $u$ and $v$ is characterized by two colors, one on the $u$ side and another on the $v$ side. More in detail, the framework defines: (i) a colour space $[1, \dots, C]$; (ii) a colored degree distribution $p$, where $p(x_1, \dots, x_C)$ denotes the probability for a vertex to have $x_i$ adjacent edges of colour $\textit{i}$; (iii) a real, symmetric $C \times C$ matrix $\textbf{T} = \{T_{ij} \geq 0\}$, called \emph{color preference matrix} and indicating the relative abundance of edges on color pairs $i,j$. The matrix is normalized, so that $\forall i \ \sum_{j=1}^{C} T_{ij} {\langle x_j \rangle}=1$, where $\langle x_j \rangle$ indicates the mean of the degree distribution on color $j$.

The mapping from a set of interdependent networks to a CDRG is straightforward. First we put all the vertexes from the different networks together. We remind the reader that the vertex sets in different networks are disjoint. Then, we associate a different color to each network and to each pair of networks. As an example, a two-layer interdependent network $G$ is mapped to a graph with three edge colors, one for the edges between vertexes originally in layer 1, one for the edges between vertexes originally in layer 2, and one for edges between one layer and the other. Notice that (i) we focus on undirected edges, (ii) every edge has the same color at its two ends (no bi-colored edges exist), and (iii) in the general case a network with $L$ layers is mapped to a graph with $L + \frac{L (L-1)}{2}$ colors.

The mapping allows us to use the following result, proved in 
\cite{Soderberg2003c} for CDRGs: Let $\textbf{J}$ be the Jacobian matrix defined by $\textbf{J=TE}$, where $\textbf{E}$ is 
defined as $E_{ij}=\partial_{z_i} \partial_{z_j} H$ and $H(z_1, \dots, z_C)$ is the multivariate generating function of the colored degree distribution $p$. If the largest eigenvalue of \textbf{J}(\textbf{1}) is larger than 1, then there is a high probability that a giant component exists.

Therefore, we need to compute $\textbf{J(\textbf{1})=TE(\textbf{1})}$ on the graph obtained after the mapping. We show the details for the case of two coupled networks. In our case there are no edges between different colors, so $T$ is a diagonal matrix with $T_{ii}=1/{\langle x_i \rangle}$ (please notice that this is true for any number of networks, not just two). Moreover, let $H(z_1,z_2,z_3) = \sum_i \sum_j \sum_k p(i, j,k)z_1^iz_2^jz_3^k$ be the multivariate generating function of the degree distribution $p(x_1,x_2,x_3)$, where $x_1, x_2$ and $x_3$ indicate node degrees respectively on the two networks and on the interconnection layer. Then, the multivariable combinatorial moments can be written as:

\begin{equation}
 E_{ij}=\partial_{z_i} \partial_{z_j} H(z_1,z_2,z_3)=\sum_i i \sum_j j \sum_k p(i, j,k)z_1^{(i-1)}z_2^{(j-1)}z_3^k
\end{equation}
\begin{equation}
 E_{ii}=\partial_{z_i} \partial_{z_i} H(z_1,z_2,z_3)=\sum_i i(i-1) \sum_j \sum_k p(i, j,k)z_1^{(i-2)}z_2^jz_3^k
\end{equation}

Assuming that the degree distributions on different colors are probabilistically independent, we can conclude that at $z_1=1, z_2=1, z_3=1$ we have:
\begin{equation}
 E_{ij}(1,1,1)=\partial_{z_i} \partial_{z_j} H(1,1,1)=\langle x_i (x_i-1) \rangle
\end{equation}
\begin{equation}
 E_{ii}(1,1,1)=\partial_{z_i} \partial_{z_i} H(1,1,1)=\langle x_i^2 - x_i \rangle
\end{equation}

We can finally write \textbf{J} computed at point (\textbf{1}) as:
\begin{equation}\label{eq:jacobOrginal}
\textbf{J(\textbf{1})=TE(\textbf{1})=}
\begin{pmatrix}
  {{\langle x_1^2 - x_1 \rangle}\over{{\langle x_1 \rangle}}} &
  {{\langle x_1 x_2\rangle} \over {\langle x_1 \rangle}} &
  {{\langle x_1 x_3 \rangle} \over {\langle x_1 \rangle}}\\
  {{\langle x_2 x_1 \rangle} \over {\langle x_2 \rangle}}  & 
  {{\langle x_2^2 - x_2 \rangle}\over{\langle x_2 \rangle}}&
 {{\langle x_2 x_3 \rangle} \over {\langle x_2 \rangle}} \\
   {{\langle x_3 x_1 \rangle} \over {\langle x_3 \rangle}}  & 
  {{\langle x_3 x_2 \rangle} \over {\langle x_3 \rangle}}&
  {{\langle x_3^2 - x_3 \rangle}\over{{\langle x_3 \rangle}}} &
 \end{pmatrix}
 \end{equation}  


Now we want to express the same matrix using the original degree distributions on the input networks, which is not the same as in the colored graph. For example, in Figure \ref{fig:kindsof}(a) the degree distribution on color $x_1$, i.e, only considering edges in $L_1$, is: $\{1,3,2,3,2,1,0,0,0,0,0,0\}$. The nodes in $L_2$ also contribute to this distribution, each with a 0. To express the condition for an epidemic based on the structure of the original networks we want to rewrite $x_1$ using the original degree distribution on $L_1$, that is, $\{1,3,2,3,2,1\}$, only considering nodes in $L_1$. Notice that this distribution, that we call $y_1$, has a higher average than $x_1$.

More in general, let $n_i$ and $n$ be respectively the number of nodes in layer $i$ and the number of nodes in the entire network ($n={\sum_{i=[1..C]} n_i}$, where $C$ is number of colors in the CDRG framework and $C = L + \frac{L (L-1)}{2}$ where $L$ is the number of networks).
Then, for $i \neq j$ we have:
\begin{equation}\label{eq:g11}
{\langle {x_i} \rangle} ={{n_i}\over{n}} {\langle {y_i} \rangle} 
\end{equation}
\begin{equation}\label{eq:g22}
{\langle {x_i x_j} \rangle} = {{n_i n_j}\over{n^2}} {\langle {y_i y_j} \rangle} 
\end{equation}
\begin{equation}\label{eq:g33}
{{\langle x_i^2 - x_i \rangle}}  =
{({{n_i}\over{n}})^2}
{{\langle y_i^2 - y_i \rangle}}
\end{equation}
Since $n_3=n_1+n_2=n$, we can rewrite the Jacobian matrix as
\begin{equation}\label{eq:jacobOrginal3}
\textbf{J}(\textbf{1})=
\begin{pmatrix}
  ({{n_1}\over{n}}) {{\langle y_1^2 - y_1 \rangle}\over{{\langle y_1 \rangle}}} &
  ({{n_2}\over{n}}) {{\langle y_1 y_2\rangle} \over {\langle y_1 \rangle}} &
   {{\langle y_1 y_3 \rangle} \over {\langle y_1 \rangle}}\\
  ({{n_1}\over{n}}) {{\langle y_2 y_1 \rangle} \over {\langle y_2 \rangle}}  & 
  ({{n_2}\over{n}}) {{\langle y_2^2 - y_2 \rangle}\over{\langle y_2 \rangle}}&
 {{\langle y_2 y_3 \rangle} \over {\langle y_2 \rangle}} \\
  ({{n_1}\over{n}}) {{\langle y_3 y_1 \rangle} \over {\langle y_3 \rangle}}  & 
  ({{n_2}\over{n}}) {{\langle y_3 y_2 \rangle} \over {\langle y_3 \rangle}}&
  {{\langle y_3^2 - y_3 \rangle}\over{{\langle y_3 \rangle}}} 
 \end{pmatrix}
 \end{equation}  

The matrix we have obtained so far corresponds to the whole colored graph, so the aforementioned condition on its largest eigenvalue only tells us about the existence of a giant component in the network. Now we are ready to study what happens on the diffusion network, that is, when only some of the links are used to propagate an infected status according to diffusion rates $\beta_1$, $\beta_2$, and $\beta_3$ (for intra- and inter-layer links). In practice, we want to substitute the degree distributions $y_i$ with new distributions $\overline{y}_i^{\prime}$ obtained by removing some links.

Along the
same line as in \cite{Qian2012}, we maintain the occupied links in each layer by removing the type-1, type-2, and type-3 links. As said in Section \ref{sec:pre}, the transmissibility for each of these kinds of links is defined as: 
\begin{equation}\label{eq:transLayeri}
R_i=1-{(1- \beta_i)}^\tau
\end{equation}
where $\tau$ is recovery time. As shown in \cite{Newman2002, Qian2012}, the probability generating function $\overline{g}_i$ of $\overline{y}_i$ is obtained from the probability generating function $g_i$ of $y_i$ as follows: 
\begin{equation}\label{eq:gnprim}
\overline{g}_i(z)=g_i(1+R_i(z-1))
\end{equation}
From these probability generating function we can finally obtain the averages of $\overline{y}_i$ and $\overline{y}_i^2$ to replace the corresponding values in Eq.~(\ref{eq:jacobOrginal3}):
\begin{equation}\label{eq:g1}
{\langle \overline{y}_i \rangle} = {R_i \langle y_i \rangle}
\end{equation}
\begin{equation}\label{eq:g2}
{\langle \overline{y}_i^2 \rangle} ={R_i}^2 (\langle {y_i}^2 \rangle - \langle {y_i} \rangle) + R_i {\langle {y_i} \rangle}  
\end{equation}
We thus obtain:
\begin{equation}\label{eq:jacobTrans1}
\textbf{J}(\textbf{1})=
\begin{pmatrix}
  ({{n_1}\over{n}}) R_1{{\langle y_1^2 \rangle - \langle y_1 \rangle}\over{{\langle y_1 \rangle}}} &
  ({{n_2}\over{n}}) R_2{{\langle y_2 \rangle}} &
   R_3{{\langle y_3 \rangle}}\\
  ({{n_1}\over{n}}) R_1{{\langle y_1 \rangle}}  & 
  ({{n_2}\over{n}}) R_2{{\langle y_2^2 \rangle - \langle y_2 \rangle}\over{{\langle y_2 \rangle}}} &
  R_3{{\langle y_3 \rangle}}\\
  ({{n_1}\over{n}}) R_1{{\langle y_1 \rangle}}  & 
  ({{n_2}\over{n}}) R_2{{\langle y_2 \rangle}} &
  R_3{{\langle y_3^2 \rangle - \langle y_3 \rangle}\over{{\langle y_3 \rangle}}}\\
 \end{pmatrix}
 \end{equation} 

In summary, if the largest eigenvalue $\theta$ of the Jacobian matrix of Eq.~(\ref{eq:jacobTrans1}) is larger than unity, $\theta > 1$, with high probability there exists a giant component in the diffusion network.

Given this result, the pseudocode to compute a multidimensional epidemic threshold is presented in Algorithm \ref{algo}. 
Given a disjoint interdependent network $G$ mapped to $n$ colors (i.e., individual networks and interconnection layers), the output of this algorithm is a set $M$ of tuples $(\beta_1, \beta_2,..., \beta_n)$ which constitute the multidimensional epidemic threshold. Notice that the algorithm cannot be directly executed, because the set N contains an infinite number of tuples, but it is sufficient to sample N to obtain an approximate multidimensional threshold of arbitrarily high precision. Other algorithms can be used to explore the solution space in a more efficient way, but this lies outside the scope of this article. 

\begin{pseudocode}{MultiEpiThr}{G}  \label{algo}
M=\{\}
\\
N= [0,1]^n
\\
\FOREACH t' \in N \DO
\BEGIN
\IF \nexists t \in M \mbox{ such that } t' \mbox{ dominates } t \THEN
\BEGIN
\IF \theta \geqslant 1 \THEN M \cup \{ t' \}
\END
\END
\\
\RETURN{M}
\end{pseudocode}

The above results have been obtained for an interdependent network with arbitrary degree distributions. 
Next we address two following specific cases.

\textbf{Case (i) - ER-ER-ER:} First, we assume that all the layers are generating using an Erd\H{o}s-Renyi (ER) model with random Poissonian degree distributions. For a random Poissonian degree distribution with average degree ${\langle y \rangle}$, we know that its second moment ${\langle y^2 \rangle}$ equals ${\langle y \rangle}({\langle y \rangle}+1)$. 
Then, we can rewrite the Jacobian matrix in Eq.~(\ref{eq:jacobTrans1}) using the following equation:
\begin{equation}\label{eq:MomentsER}
{{{\langle y_i^2 \rangle}-{\langle y_i \rangle}} \over {\langle y_i \rangle}}= {\langle y_i \rangle}
\end{equation}

\textbf{Case (ii) - SF-ER-SF:} Here, we assume to have two scale-free networks coupled with an ER inter-layer network. The degree distribution of both scale-free networks follows a power-law with exponent $\gamma$:
\begin{equation}\label{eq:powerLaw}
p(y)=cy^{-\gamma}, \hspace{2mm} Y_{min}\leq y \leq Y_{max}
\end{equation}
where $c=(\gamma - 1)Y_{min}^{\gamma -1}$ is a normalization constant, $Y_{min}$ is the smallest possible connectivity, and the maximum degree $Y_{max}$, called the natural upper cut-off, is chosen so that there is at most one node whose degree is higher than $Y_{max}$. The \textit{m}-th moment of such power-law degree distribution can be written as:
\begin{equation}\label{eq:Moments}
{\langle y^n \rangle}=\int_{Y_{min}}^{Y_{max}} y^np(y)dy=c {{{Y_{max}^{n-\gamma +1}} - {Y_{min}^{n-\gamma +1}}} \over {n-\gamma +1}}
\end{equation}
Then, to rewrite the Jacobian matrix for this case, we use following equations (for $i=\{1,2\}$) in the Jacobian matrix of Eq. (\ref{eq:jacobTrans1}):
\begin{equation}\label{eq:Moments2}
{\langle y_i \rangle}=(\gamma_i - 1) {Y_{min}^{\gamma_i -1}} {{{Y_{max}^{2-\gamma_i}} - {Y_{min}^{2-\gamma_i}}} \over {2-\gamma_i}}
\end{equation}
\begin{equation}\label{eq:Moments21}
{{{\langle y_i^2 \rangle}-{\langle y_i \rangle}} \over {\langle y_i \rangle}}= [({{2-\gamma_i} \over {3- \gamma_i}}) {{{Y_{max}^{3-\gamma_i}} - {Y_{min}^{3-\gamma_i}}} \over {{Y_{max}^{2-\gamma_i}} - {Y_{min}^{2-\gamma_i}}}}] -1
\end{equation}
For the interconnection layer ($i=3$) we use Eq.~(\ref{eq:MomentsER}).

\section{Simulation Results}\label{sec:results}

To verify the analytical findings obtained in the previous section, here we present our simulation results. In the simulations we consider both synthetic and real disjoint interdependent networks. We execute an SIR model with $0 < \beta \leq 1$ and $0 < \alpha \leq 1$ as intra- and inter-layer information diffusion rates, respectively. The recovery time is set to $5$ steps.
The simulation results are averaged over $100$ realizations.

\subsection{Networks}\label{subsec:networks}

In the experiments we have used the following kinds of networks.

\textbf{Erd\H{o}s-Renyi:}
In these experiments we generated two interdependent networks using the \emph{Erd\H{o}s-Renyi} model for both the networks and their interconnections \cite{Erdos1959}. $n$ labeled nodes are connected with $m$ randomly placed links. In the experiments we set $n =10000$ and added a varying number $m$ of links to obtain different average degrees (computed as $\langle x \rangle = 2m/n$).

\textbf{Scale-free:}
For most real scale-free networks the degree exponent $\gamma$ is between 2 and 3 \cite{Clauset2009}, therefore for these experiments we generated two random graphs with $1500$ nodes and power-law degree exponents $\gamma_1=2.9$ and $\gamma_2=2.1$ for layer 1 and 2, respectively. Since the average degree is larger for a smaller exponent, layer 2 is always denser and has a lower epidemic threshold.

\textbf{Real-world:}
We use the following datasets as real interdependent networks:

(i) Polblog \cite{Adamic2005}: This is a two-layer network of hyperlinks between the political weblogs related to different communities: (1) liberal (as layer 1) with 759 nodes and $7303$ links (2) conservative (as layer 2) with 735 nodes and $7841$. There are $1575$ inter-layer links. 

(ii) Flickr \cite{Wang2012}: It includes $35313$ users and $3017530$ friendship links between them, with their joined groups. We selected two groups (from 200 existing groups) of size 518 and 521 (as layers 1 and 2, respectively) and the links between them (as interconnection layer). The number of intra-layer links is $6340$ for layer 1 and $18051$ for layer 2, and there are 343 inter-layer links. 

\subsection{Main outcomes}\label{InfectionDensity}

\begin{figure*}[!htp]
\centering
{
\subfigure[
\emph{ER-ER-ER: $L_1L_2$/Strongly}
]{\includegraphics[scale=0.56]{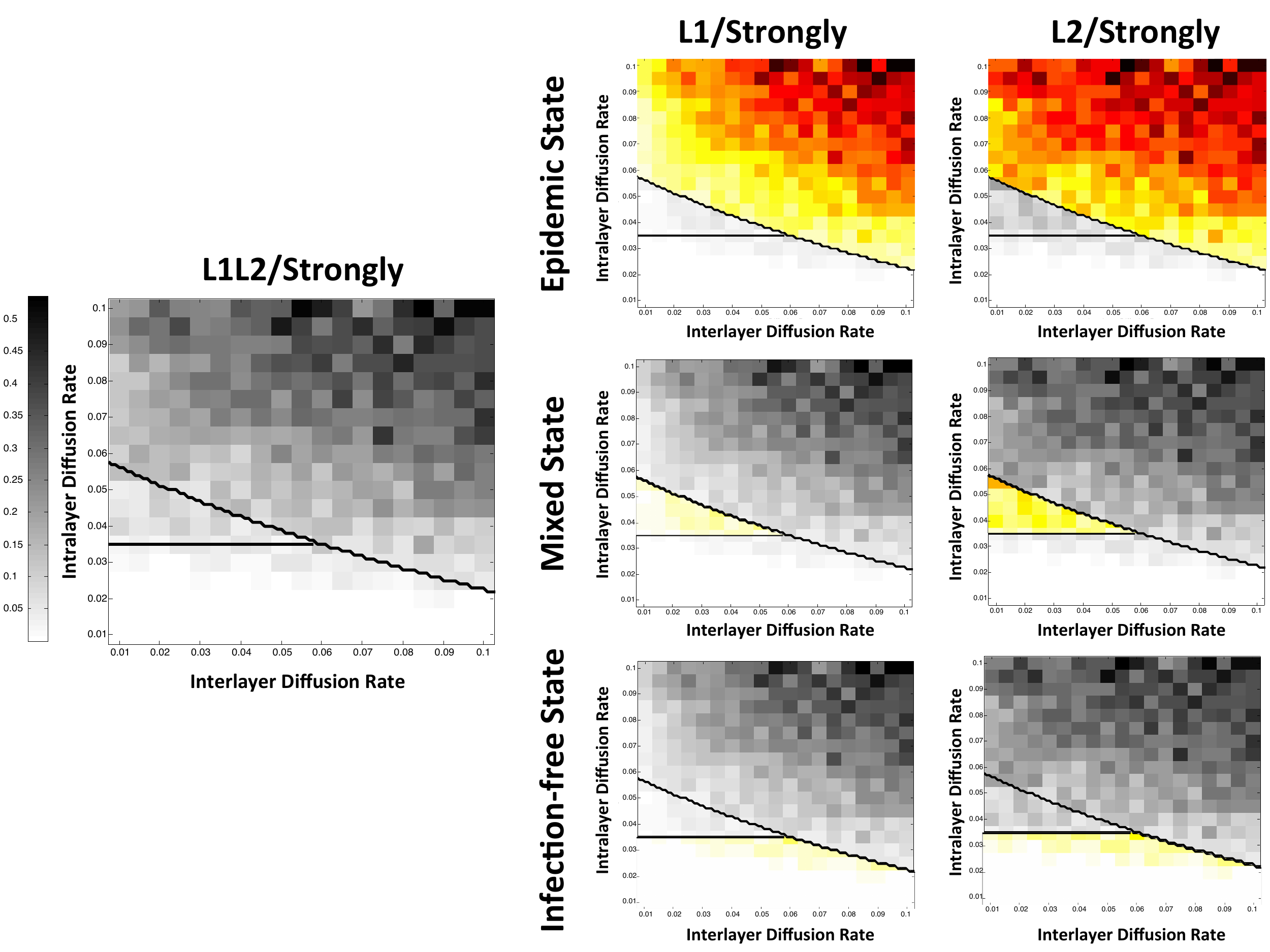}
} 
\\
\subfigure[
\emph{ER-ER-ER: $L_1L_2$/Weakly}
]{\includegraphics[trim = 10mm 65mm 1mm 60mm, clip, scale=0.35]{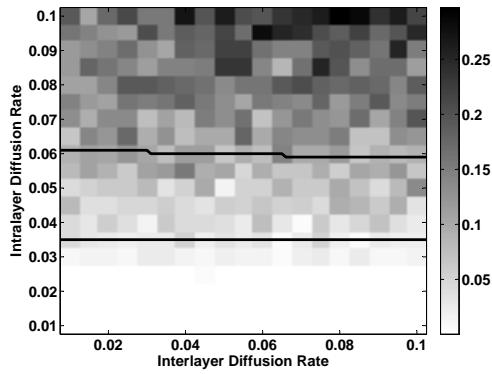}
} 
\subfigure[
\emph{SF-ER-SF}
]{\includegraphics[trim = 10mm 65mm 1mm 60mm, clip, scale=0.35]{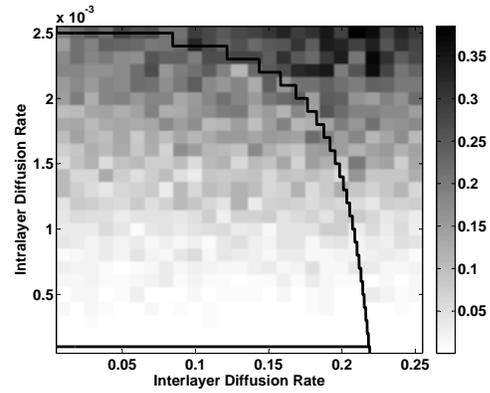}
\label{p2129e6All}
}
\subfigure[
\emph{PolBlogs}
]{\includegraphics[trim = 10mm 65mm 1mm 60mm, clip, scale=0.35]{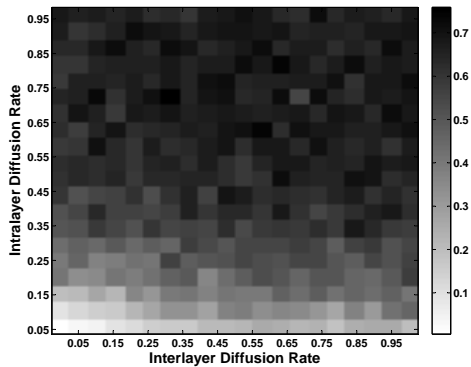}
} 
\subfigure[
\emph{Flickr}
]{\includegraphics[trim = 10mm 65mm 1mm 60mm, clip, scale=0.35]{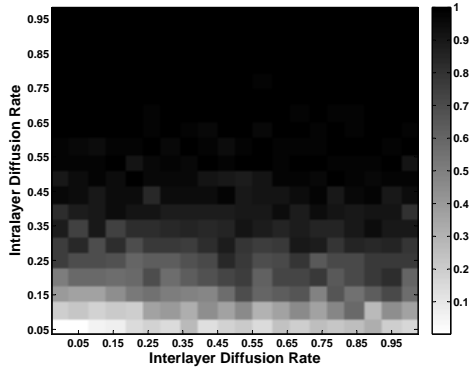}
}
}
\caption{
Heat maps of infection densities recovered from simulation results against the region boundaries derived analytically. Please, notice the different scales used in different figures: (a-b-c) present a close view of the multidimensional threshold, and higher values of X and Y are not shown as not interesting (as expected, infection density quickly increases to high values). 
(a-b) ER networks: using $\langle y_1 \rangle = 1.5, \langle y_2 \rangle = 6.0$, we show results obtained for different values of intra-layer (Y axis) and inter-layer (X axis) diffusion rates in the two cases of weakly-coupled ($\langle y_3 \rangle = 0.1$) and strongly-coupled networks ($\langle y_3 \rangle = 1.5$). 
Plot (a) also shows different states (infection-free, mixed and epidemic) in each layers of strongly-coupled case.
The number of nodes in each layer is $10000$.
(c) Scale-free networks: number of nodes in each layer = $1500$; power-law exponents: $\gamma_1=2.9$ and $\gamma_2=2.1$; average degree of inter-layer links $\langle y_3 \rangle=6.0$. 
(d-e) Real networks.}
\label{fig:density}
\end{figure*}

First, we study the infection density, i.e., the ratio of the number of infected nodes in each layer (or the entire network) to the size of giant connected component in that underlying layer (or entire network). 
For simplicity, we assume $\beta=\beta_1=\beta_2$ and $\alpha=\beta_3$ as intra- and inter-layer diffusion rates.
In Figure \ref{fig:density} we have reported some selected simulation results (as heat maps) obtained for different values of $\beta$ and $\alpha$ in which darker colors indicate higher infection density. 
Along the same line as in \cite{Dickison2012a}, we consider weakly- and strongly-coupled cases in the experiments on ER networks, 
for example $L_1$/Strongly shows the diffusion dynamic in layer 1 of a strongly-coupled network or $L_1L_2$/Strongly represents the results for entire network in a strongly-coupled case\footnote{According to Eq.~(3) in \cite{Dickison2012a}, for $\langle y_1 \rangle = 1.5$ and $\langle y_2 \rangle = 6.0$ in the case of ER-ER-ER, $\langle y_3 \rangle > 1.23$ generates a strongly connected network and $\langle y_3 \rangle < 1.23$ leads to a weakly-coupled case.}.

To compare between analytical calculations and simulation results, the region boundaries of different states (infection-free, mixed and epidemic) derived analytically (for entire network in ER-ER-ER and SF-ER-SF scenarios) are displayed in this figure against heat maps. Moreover, Figure 3(a) shows with more details the infection densities in each layer of strongly-coupled case (similar results have been obtained for other networks).
The lower straight boundary line shows the epidemic threshold for layer $2$, which is the layer with lower epidemic threshold (the epidemic threshold for an individual layer can be computed by Eq.~(\ref{eq:epiTheSingle})). 
The region below this line demonstrates the infection-free state where the infection remains limited to the neighborhood of the initial infected nodes.
Moreover, the upper boundary line represents the two-dimensional epidemic threshold (It can be calculated by Algorithm \ref{algo}). The region above this line shows the epidemic state where the infection spreads throughout the entire network.
Furthermore, the region between these lines is the mixed state where the infection can be seen to become epidemic only in layer $2$.

Comparing these region boundaries and the intensity of colors in heat maps shows agreement between the patterns observed in analytical studies and numerical simulations. 
For example, as we can see in Figure 3(a), the color of the heat map in the region corresponding to the mixed state of $L_2$ is darker than for $L_1$. 
This shows that the infection first enters the epidemic phase in this region and diffuses across $L_2$ while the infection density in $L_1$ remains negligible. Moreover, above the two-dimensional epidemic threshold the infection density in $L_1$ increases, showing that for this region the entire network is in the epidemic state.

Moreover, a comparison of the strongly- and weakly-coupled cases (Figures 3(a) and 3(b), respectively) shows how increasing the coupling strength (i.e., $\langle y_3 \rangle$) makes the role of the inter-layer diffusion rate more important in determining the epidemic threshold of the entire network. This also decreases the epidemic thresholds of the layers and consequently the area of mixed state. However, we can observe a mixed state even for strongly connected networks. 

Furthermore, for the scale-free network in Figure 3(c) and the real datasets  in Figure 3(d-e) the epidemic threshold is close to 0: most of the points have dark colors, indicating an epidemic state. As expected from the theory, our boundary lines show that the infection-free state for scale-free networks is very small. In Figure 3(c) we can see a small discrepancy between the simulation results and the expected boundary. While this difference is emphasized by the large scale used for the X and Y axes and practically affects only a small range of input values (note that the actual scale of these axes are proportional to $10^{-3}$).

In addition, we find that a smaller difference between the epidemic thresholds of the coupled layers leads to a smaller area where a mixed state occurs. This shows that the appearance of a mixed state depends either on the coupling strength between layers or on the structure of the layers (in particular, the difference between the layers' epidemic thresholds). Therefore, there is no mixed state in an interdependent network where the different layers are equal (i.e., there is no difference between the epidemic thresholds in these layers) for both weakly- and strongly-coupled cases.  

\begin{figure*}[!htp]
\centering
\subfigure[
\emph{$L_1$/Weakly}
]{\includegraphics[trim = 3mm 60mm 1mm 33mm, clip, scale=0.28]{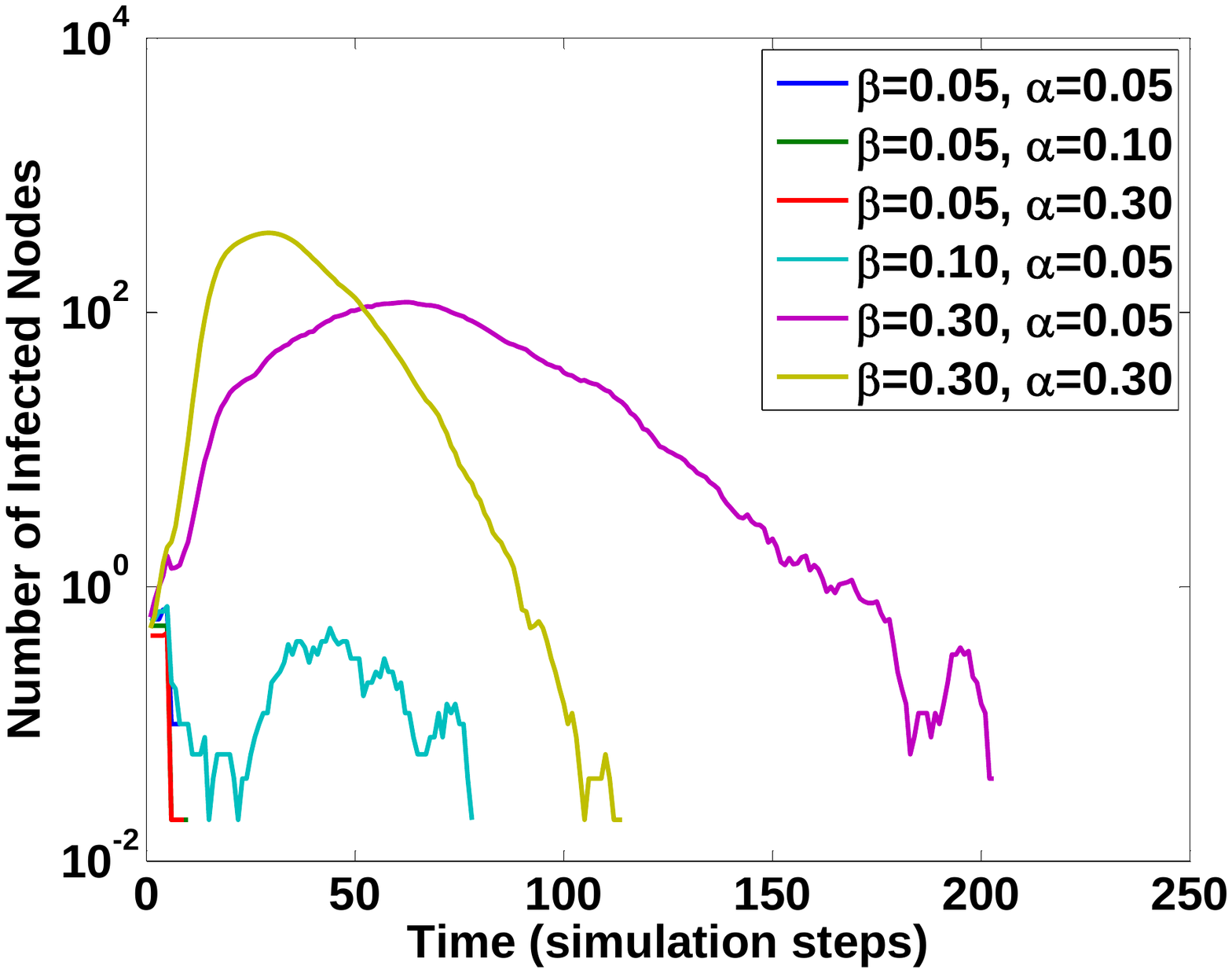}
\label{fig:Sc2-weeklyLayer0}}
\subfigure[
\emph{$L_2$/Weakly}
]{\includegraphics[trim = 3mm 60mm 1mm 33mm, clip, scale=0.28]{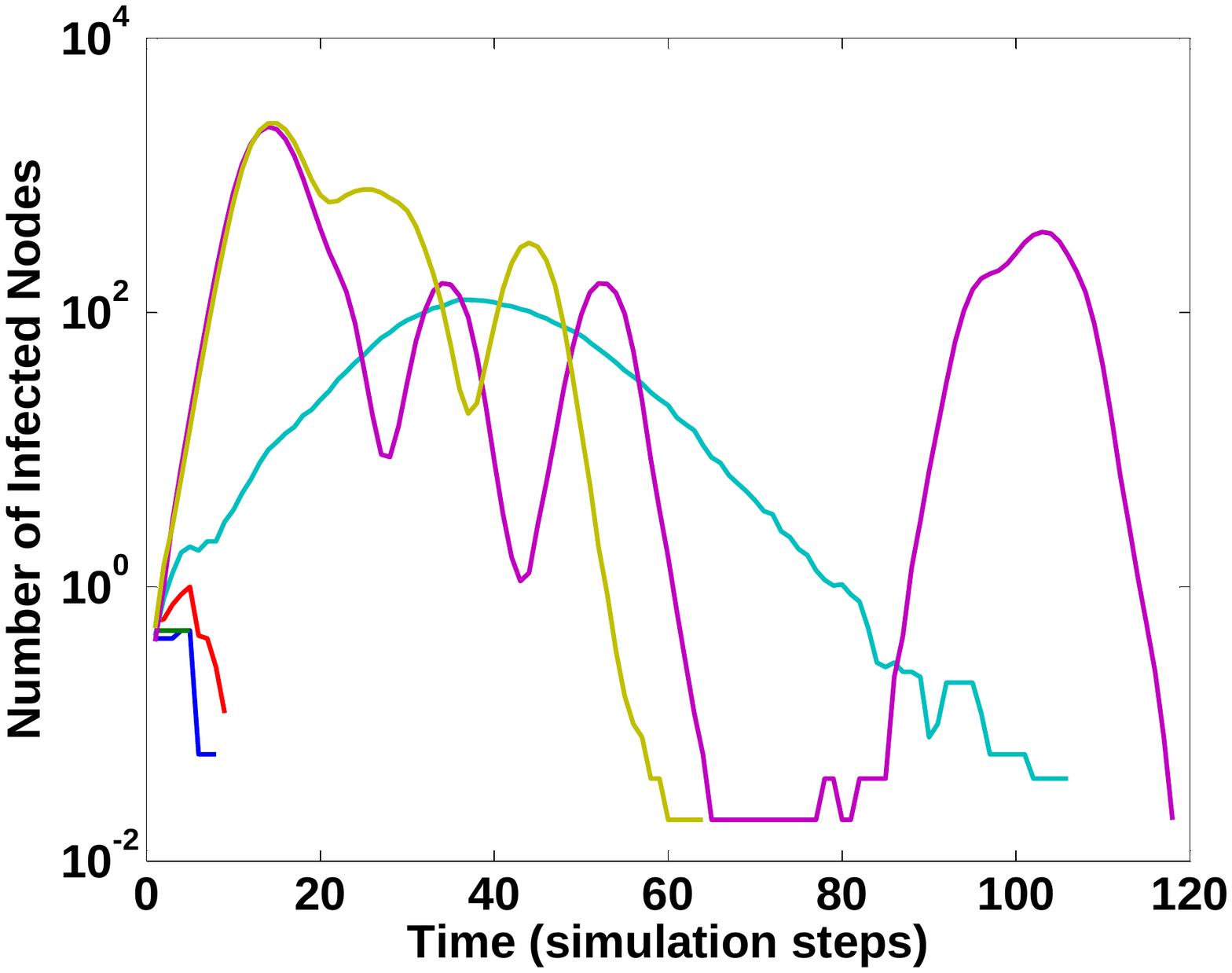}
\label{fig:Sc2-weeklyLayer1}}
\subfigure[
\emph{$L_1$/Strongly}
]{\includegraphics[trim = 3mm 60mm 1mm 33mm, clip, scale=0.28]{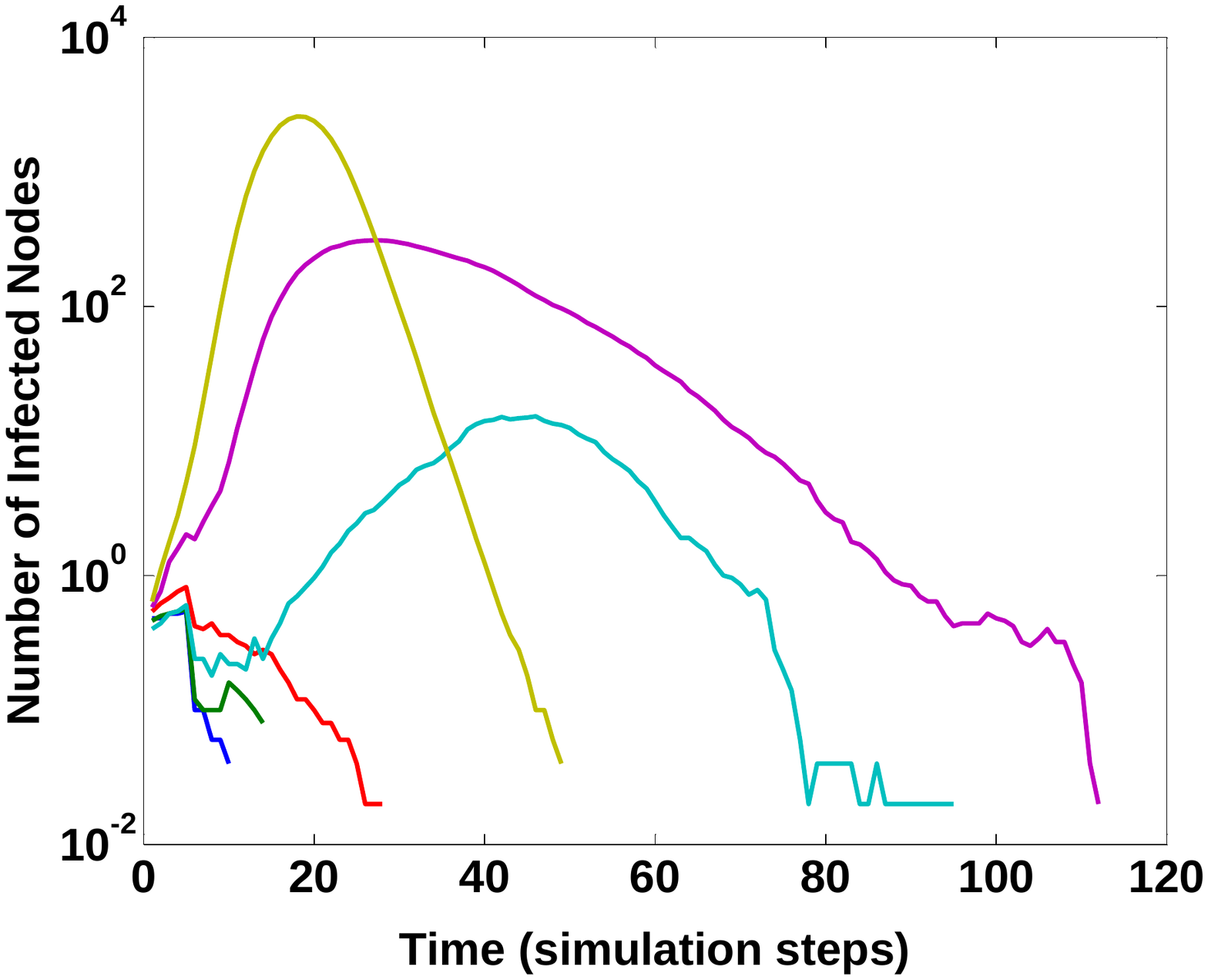}
\label{fig:Sc2-stronglyLayer0}}
\subfigure[
\emph{$L_2$/Strongly}
]{\includegraphics[trim = 3mm 60mm 1mm 33mm, clip, scale=0.28]{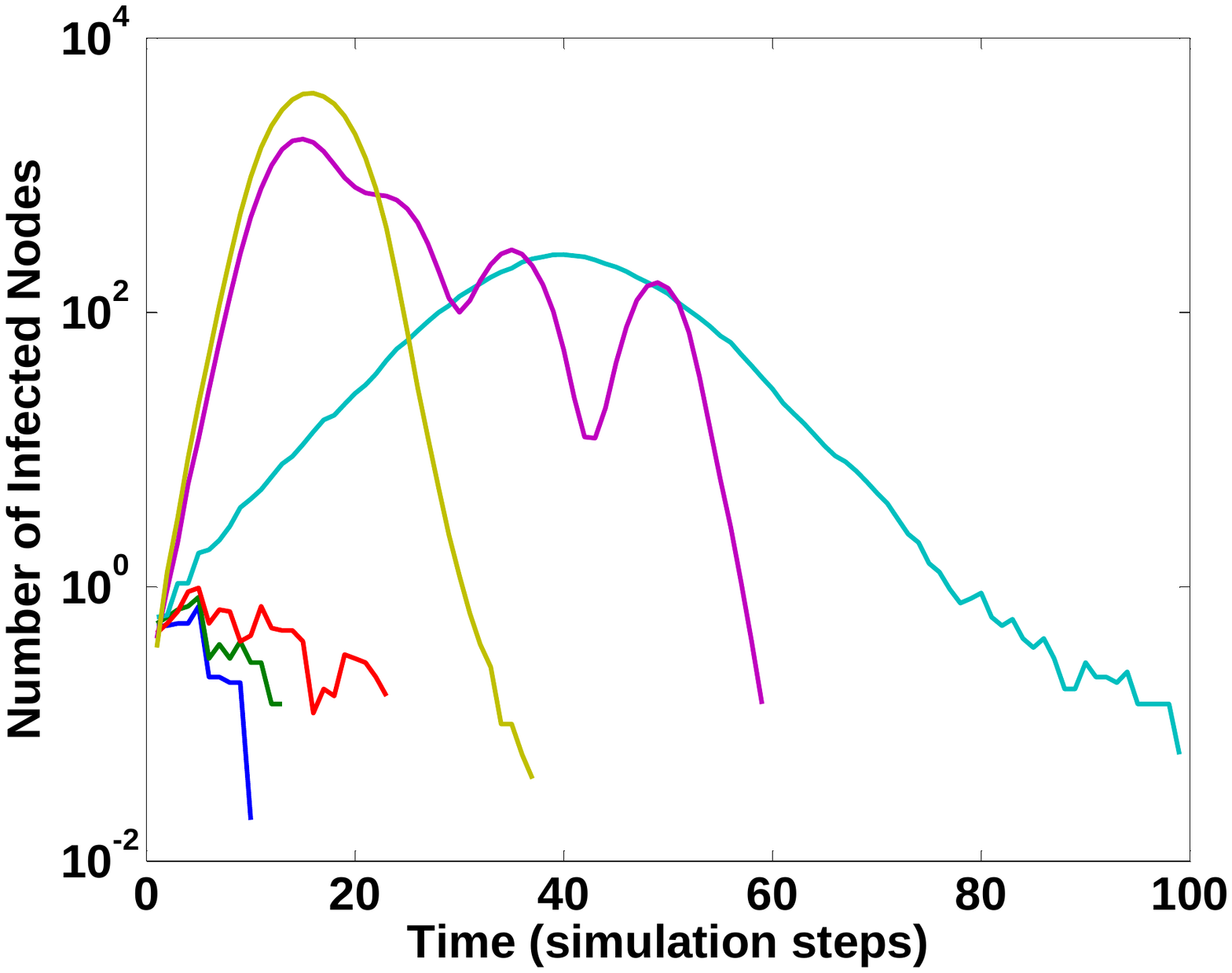}
\label{fig:Sc2-stronglyLayer1}} 
\subfigure[
\emph{$L_1L_2$/Weakly}
]{\includegraphics[trim = 3mm 60mm 1mm 33mm, clip, scale=0.28]{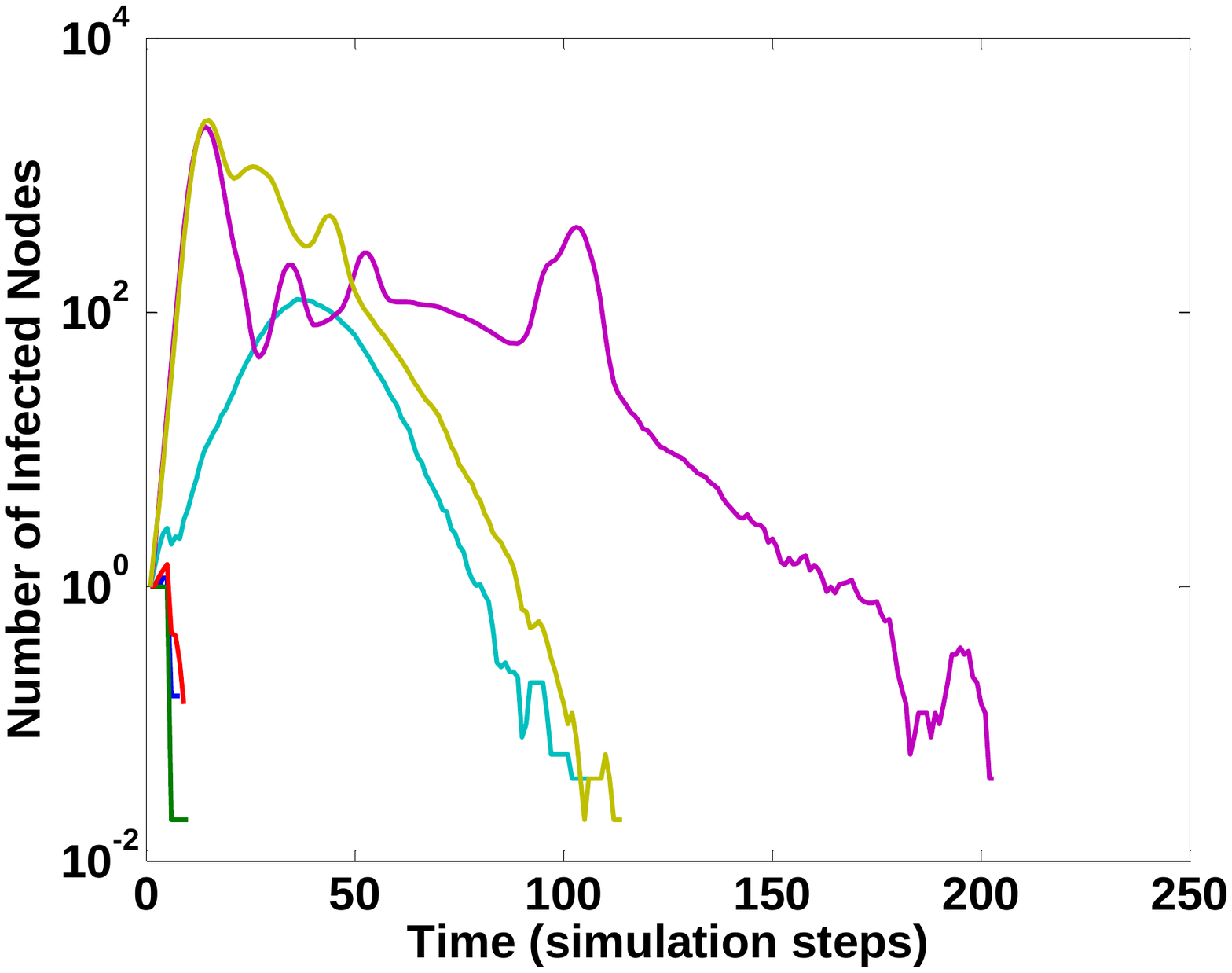}
\label{fig:Sc2-weeklyAll}} 
\subfigure[
\emph{$L_1L_2$/Strongly}
]{\includegraphics[trim = 3mm 60mm 1mm 33mm, clip, scale=0.28]{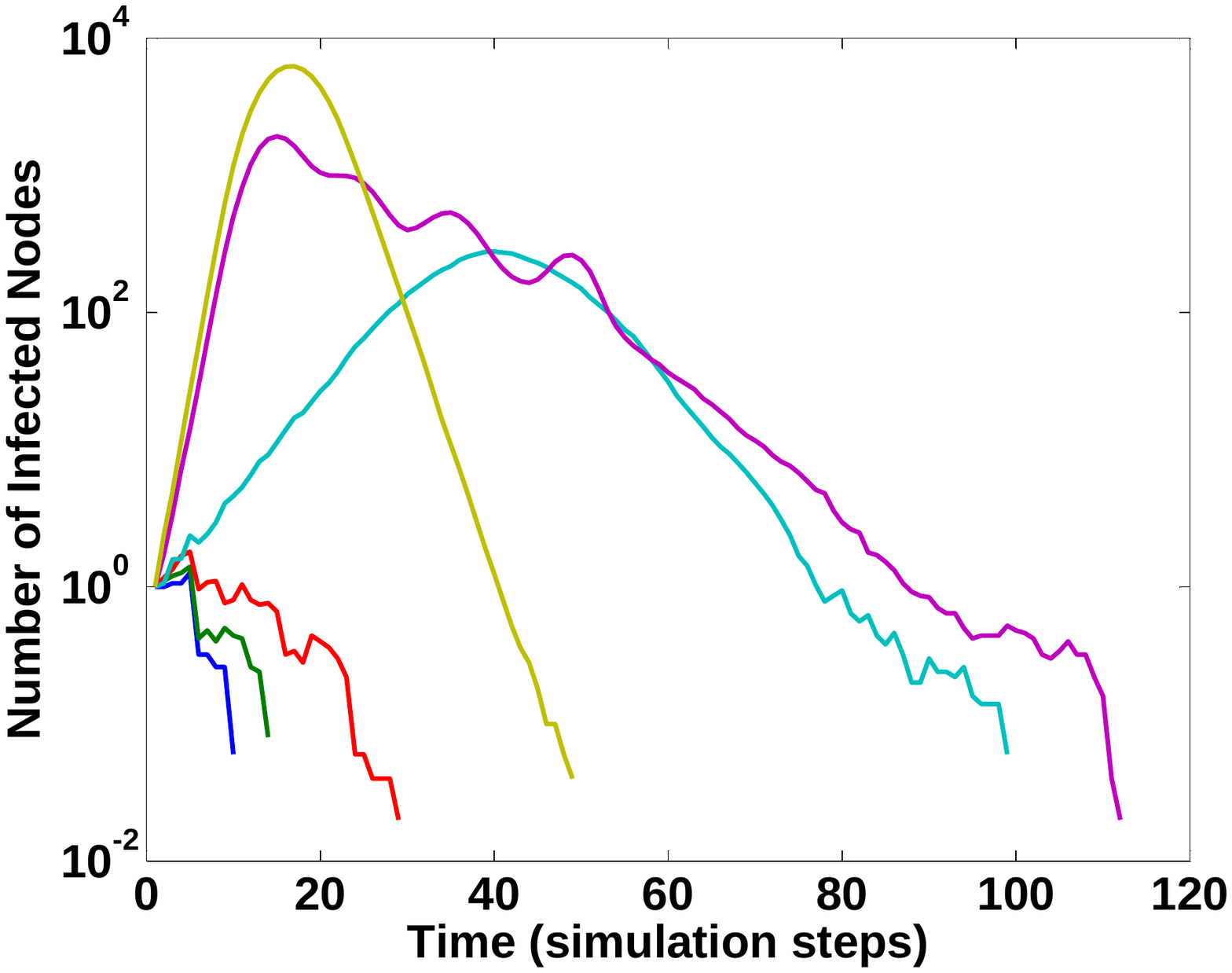}
\label{fig:Sc2-stronglyAll}} 
\subfigure[
\emph{$L_2$/SF-ER-SF}
]{\includegraphics[trim = 3mm 60mm 1mm 33mm, clip, scale=0.28]{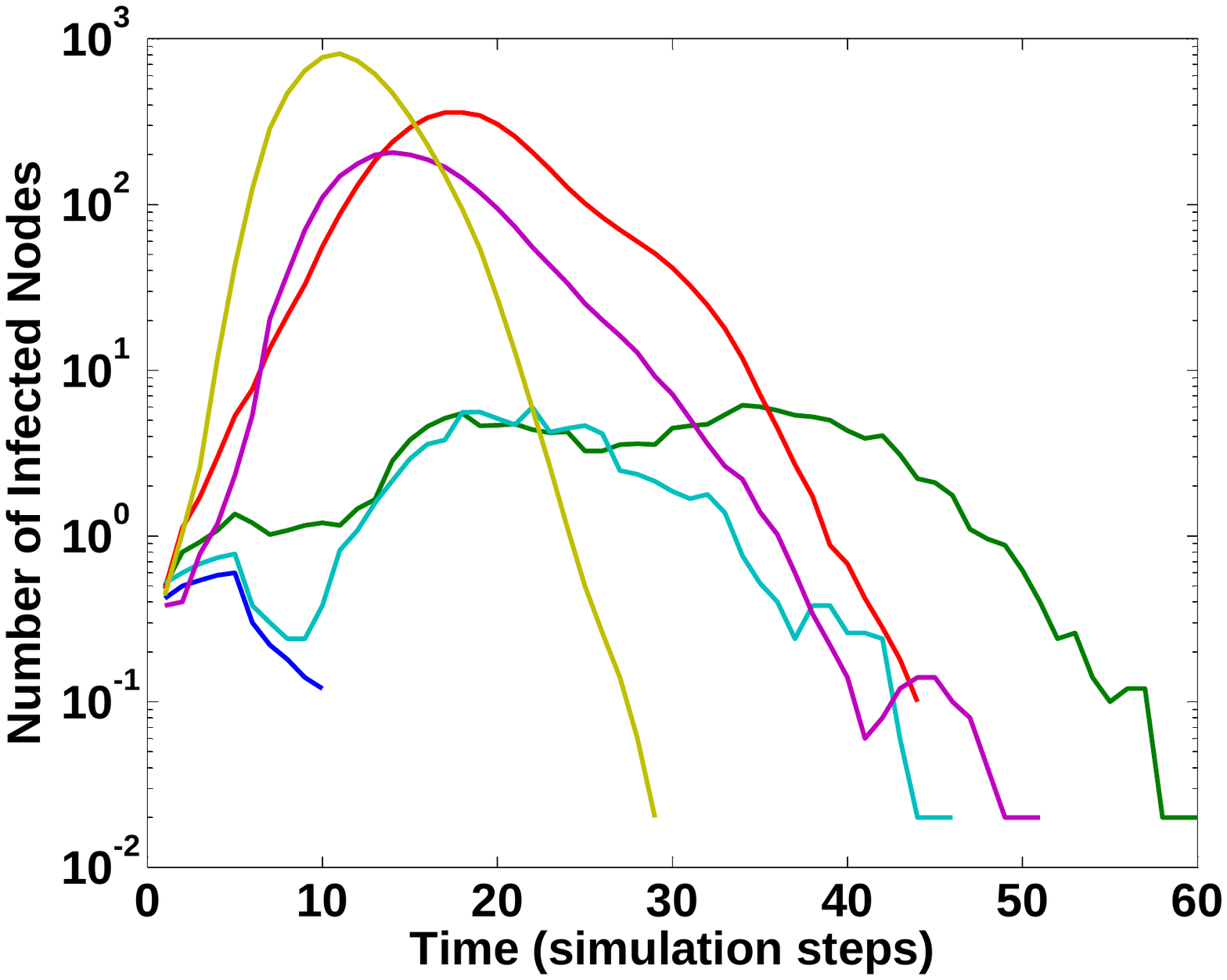}
}
\subfigure[
\emph{$L_2$/PolBlogs}
]{\includegraphics[trim = 3mm 60mm 1mm 33mm, clip, scale=0.28]{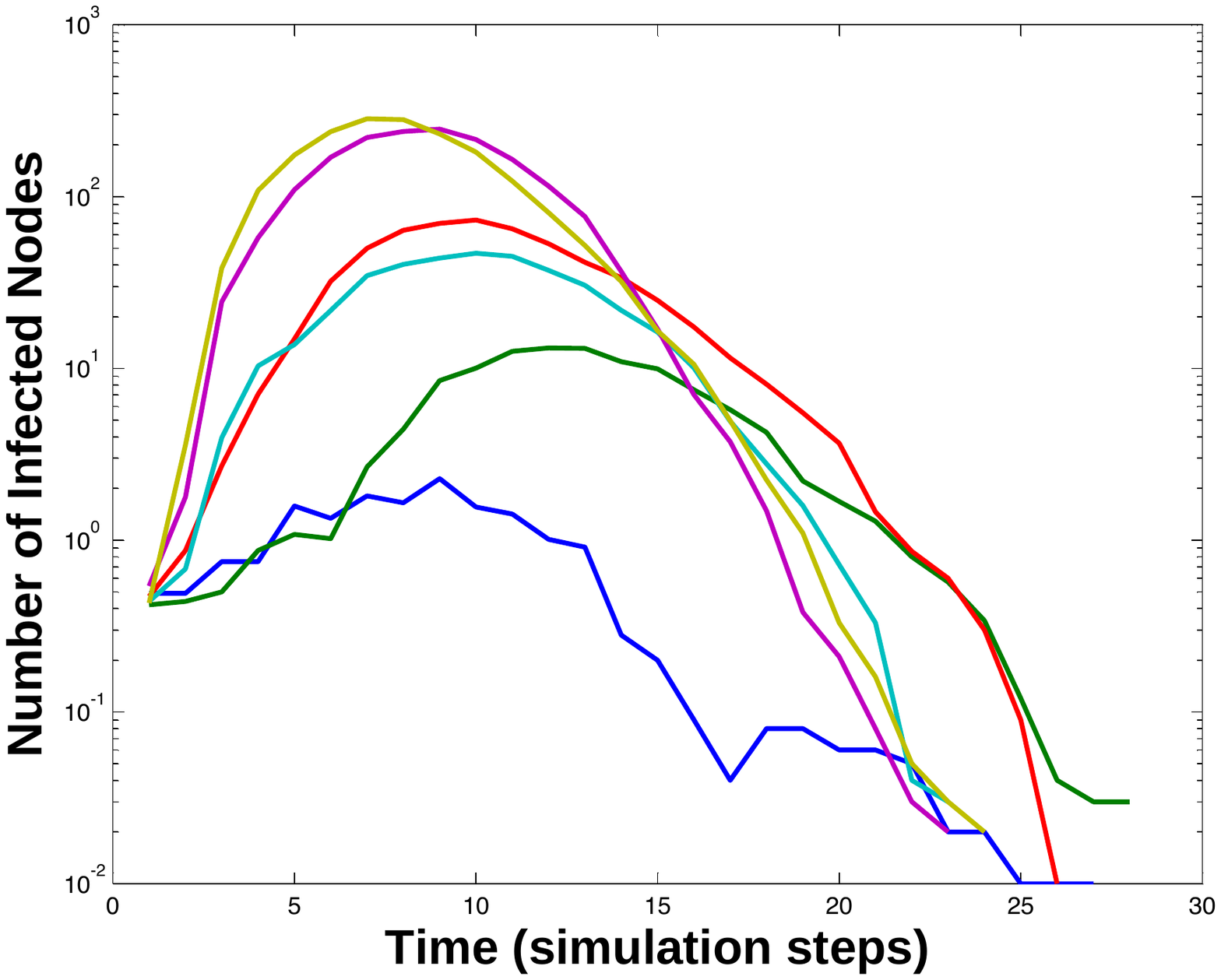}
}
\subfigure[
\emph{$L_2$/Flickr}
]{\includegraphics[trim = 3mm 60mm 1mm 33mm, clip, scale=0.28]{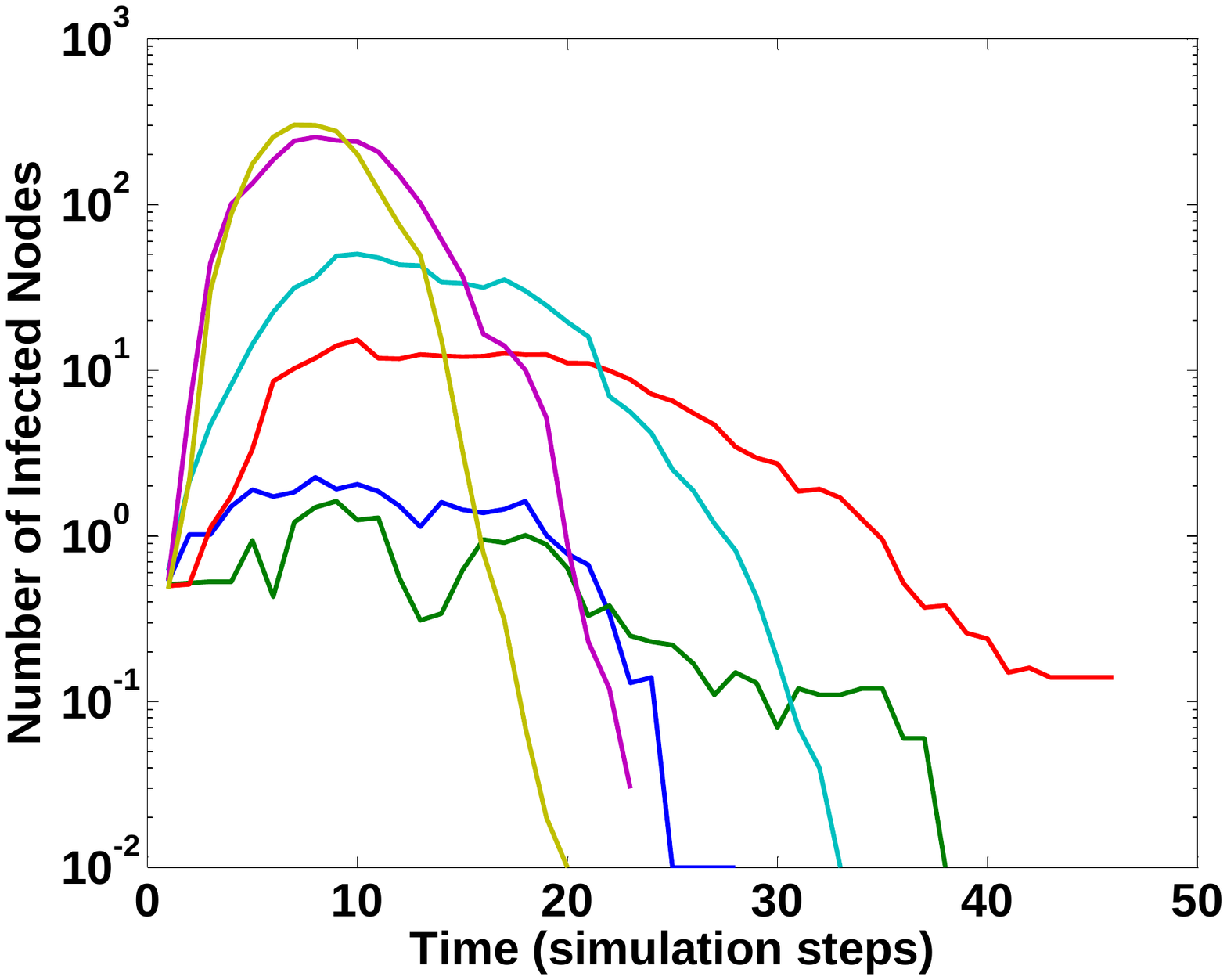}
}
\caption{Diffusion dynamics, i.e. the evolution of the number of infected nodes in time.
(a-f) ER networks: setting $\langle y_1 \rangle = 1.5$ and $\langle y_2 \rangle = 6.0$, we show the results obtained for different values of $\beta$ and $\alpha$ (as intra- and inter-layer diffusion rates, respectively) for two cases of weakly-coupled ($\langle y_3 \rangle = 0.1$) and strongly-coupled networks ($\langle y_3 \rangle = 1.5$).  
$\langle x_1 \rangle = 1.5$, $\langle x_1 \rangle = 6.0$,  and $\langle x_3 \rangle < 1.23$ indicate the weakly-coupled case studied in \cite{Dickison2012a}. The number of nodes in each layer is $10000$.
(g) Scale-free networks: number of nodes in each layer = $1500$; power-law exponents: $\gamma_1=2.9$ and $\gamma_2=2.1$; average degree of inter-layer links $\langle y_3 \rangle = 6.0$. (h-i) Real networks.
}
\label{fig:Real-dyn}
\end{figure*}

In Figure \ref{fig:Real-dyn} we show diffusion dynamics, in particular the number of infected nodes in time. In each plot the six curves correspond to six different settings of $\beta$ and $\alpha$ --- refer to the legend in Figure 4(a).
This figure can be used to check if any of these six tuples $(\beta$, $\alpha)$ can lead to a mixed state or not. In particular, if in layer 2 (as the layer with lower epidemic threshold) we observe significant peaks for a tuple, this means that the entire network is in mixed state for that tuple. 
In the weakly-coupled case a mixed state can be observed for $(\beta=0.30, \alpha=0.05)$ and $(\beta=0.30, \alpha=0.30)$.
This result is in line with the one obtained in Figure \ref{fig:density}. 

The region where a mixed state is expected is instead very small when scale-free and real networks are involved, because of the very small epidemic thresholds associated to these network structures. In this case the multidimensional case is thus not significantly different from single-network behaviors, and in fact we cannot observe this state for any settings of $(\beta, \alpha)$. For scale-free and real networks we have only shown results concerning layer $2$ in Figures 4(g-i) --- similar patterns have been obtained for layer 1 and for the entire network. 

\newgeometry{left=0.3cm, right=0.3cm, top=0.3cm, bottom=0.3cm}

 \restoregeometry

\newgeometry{left=0.5cm, right=0.5cm, top=2cm, bottom=2cm}

\restoregeometry

\section{Related Work}\label{sec:relatedWork}

The study of multilayer networks is a rapidly evolving research area with many challenging research issues. For a survey on different aspects of multilayer networks we refer to \cite{Kivela2013, Boccaletti2014} and references therein. A comprehensive review of diffusion processes over multilayer networks is available in \cite{Salehi2014}. In this section we mainly focus on disjoint interdependent networks, object of this article.

In \cite{Dickison2012a} the authors address the impact of interaction (or coupling) strength between layers on the dynamics of disease spreading in an interdependent network. In this work the parameter $\kappa$ (also used in our article) is used as a measure of coupling strength. 
A two-layer interdependent network is defined as strongly-coupled if $\kappa_T$ is larger than both $\kappa_A$ and $\kappa_B$, where $\kappa_T$ is calculated over the entire coupled network and $\kappa_A$ and $\kappa_B$ are computed over the individual networks $A$ and $B$. In addition, a network is defined as weakly-coupled if $\kappa_B > \kappa_T$ and $\kappa_T > \kappa_A$. The authors find that a mixed phase can happen in a weakly-coupled network, where the disease is epidemic on only one layer (i.e., layer $A$). 
As another measure of interaction strength, the inter-layer link density (i.e., the ratio of the existing inter-layer links between two layers to the total number of possible such links) is utilized in \cite{Wang2011a} to study the effects of inter-layer links on information spreading in two-layer interdependent networks. The authors show that more inter-layer links leads to a larger number of infected nodes. Moreover, the infection peak occurs in the two layers at different times when they are sparsely interconnected. 

The impact of different diffusion rates is not considered in any of these works \cite{Dickison2012a, Wang2011a}. In \cite{Min2013}, the authors study the effect of layer-switching cost (defined as difference between intra- and inter-layer diffusion rates) on diffusion processes over multiplex networks. They find that larger differences between intra- and inter-layer diffusion rates lead to larger epidemic thresholds.
Moreover, different intra- and inter-layer diffusion rates have been considered in some recent studies on various applied scenarios \cite{Yagan2013, Qian2012, Qian2013}. 
The authors of \cite{Yagan2013} define different speeds for information spreading over each layer of a physical-social network. They show that an epidemic state can occur in the whole network even when the information does not propagate inside a particular layer. 
The authors of \cite{Qian2012} studied the impact of clique structures on the speed of information diffusion. They show that propagation of information when large cliques are present is faster. 
In \cite{Qian2013}, the same authors show that, in a social-physical network, increasing the size of the online social network could even decrease the number of infected nodes. However, the focus of all these works is on multiplex networks.

To the best of our knowledge, our work is the first to study information diffusion over disjoint interdependent networks with arbitrary degree distributions and  considering different diffusion rates.

\section{Conclusion}\label{sec:discussion}

Diffusion processes on multilayer networks are intrinsically multidimensional and studying them using a single parameter for all networks cannot provide an accurate description of real propagation events. Therefore, we studied both analytically and by simulation the relation between epidemic threshold and diffusion rates in the general case of interdependent networks with different diffusion rates. 

The existence of a multidimensional threshold tells us that the structure of the networks and of their interconnections in combination with different diffusion rates can determine alternative propagation patterns. Propagation can follow different paths, enabling the emergence of epidemics on networks that would otherwise be characterized by a very high mono-dimensional epidemic threshold, and we have shown that in theory a mixed phase can occur even when the two networks are strongly coupled by setting specific diffusion rates.

However, at the same time our extension of previous results to generic degree distributions has shown that in real cases it is unlikely to observe these complex patterns, and diffusion processes are expected to 
conquer quickly the multiple networks without showing significant differences. This result is not unexpected, and motivates the development of more realistic models to study real phenomena, as the one presented in this work. In the future, with the increasing availability of high-quality data on disjoint interdependent networks and more in general multilayer networks it will be interesting trying to identify if some of the other less likely scenarios predicted by our model can be observed in real data.

\section*{References}


\end{document}